\documentclass[preprint,12pt]{elsarticle}

\usepackage{graphicx}
\usepackage{amssymb}
\usepackage{amsmath}
\usepackage[ruled,vlined]{algorithm2e}
\usepackage{lineno}
\usepackage{hyperref}
\usepackage{aas_macros}
\usepackage{multirow}

\graphicspath{ {./images/} }

\begin{document}

\begin{frontmatter}


\title{Photometric Data-driven Classification of Type Ia Supernovae in the Open Supernova Catalog}

\author[hse]{Stanislav Dobryakov}
\ead{stdobr@gmail.com}

\author[sai,hse,uiuc]{Konstantin Malanchev}
\ead{malanchev@physics.msu.ru}

\author[hse]{Denis Derkach}
\author[hse]{Mikhail Hushchyn}

\address[hse]{National Research University Higher School of Economics\\ 20 Myasnitskaya Ulitsa, Moscow 101000, Russia}
\address[sai]{Sternberg Astronomical Institute, Lomonosov Moscow State University\\ 13 Universitetsky pr., Moscow 119234, Russia}
\address[uiuc]{Department of Astronomy, University of Illinois at Urbana--Champaign\\ 1002 West Green Street, Urbana, IL 61801, USA}

\begin{abstract}
We propose a novel approach for a machine-learning-based detection of the type Ia supernovae using photometric information. Unlike other approaches, only real observation data is used during training. Despite being trained on a relatively small sample, the method shows good results on real data from the Open Supernovae Catalog. We also investigate model transfer from the PLAsTiCC simulations train dataset to real data application, and the reverse, and find the performance significantly decreases in both cases, highlighting the existing differences between simulated and real data.
\end{abstract}


\end{frontmatter}


\section{Introduction\label{sec:intro}}
Supernovae provide an enormous amount of information for astrophysics researchers. For example, Type Ia Supernovae (SNIa) are used as standardisable candles, that makes them highly important for extragalactic astrophysics and cosmology allowing to measure distances across the Universe \cite{1974PhDT.........7R,1977SvA....21..675P,1998AJ....116.1009R,1999ApJ...517..565P,2018ApJ...859..101S,2019ApJ...872L..30A}.

Traditionally, SNIa candidates discovered photometrically are checked by spectroscopic follow-up, that requires a significant amount of dedicated observational resources. 
However, in the epoch of large synoptic surveys such as Zwicky Transient Facility\footnote{\url{http://ztf.caltech.edu}}~\cite{ztf} and the Vera~C.~Rubin Observatory Legacy Survey of Space and Time (LSST)\footnote{\url{http://lsst.org}}~\cite{lsst} hundreds of thousands of transient candidates are expected to be found per night, which makes it impossible to confirm all of them spectroscopically.
Moreover, it is not possible to perform additional observations for such short-lived historical objects.
That is why photometrically classified SNe are used in cosmological analysis, for example, SNe discovered by Foundation Supernova Survey\citep{Jones_etal2017,Jones_etal2018,Jones_etal2019}.
Thus, it has become essential to classify transients, based on photometric information alone, to have an opportunity to find suitable SNIa candidates, even in the case of a lack of spectroscopic observations.

Photometric supernovae classification, using the machine-learning approach is a well-developed field of research.
Several simulated datasets are available from The Supernova Photometric Classification Challenge (SPCC)~\cite{Kessler_etal2019} and Photometric LSST Astronomical Time-Series Classification Challenge (PLAsTiCC)~\cite{PLAsTiCC}. These datasets contain light curves of different types of transients, in the way they could be observed by the Dark Energy Survey~\footnote{\url{http://darkenergysurvey.org}} and LSST correspondingly.
These challenges, and their datasets, inspired several papers aimed at providing a solution for the photometric classification of supernovae (see, e.g. \cite{Lochner_2016,2018MNRAS.473.3969R,Markel_etal2019,Muthukrishna_etal2019,vargas-dos-santos_etal2019,Moller_Boissi2020}).
Supernova data, from these challenges, use spectroscopic templates~\cite{snana2009,salt2_2007,sugar2019}, thus every dataset corresponds to one of a few real historical supernovae observed, at a different distance and with a different cadence.
However, Type Ia supernovae are very diverse~\cite{blondin_etal2012}, and such simulations do not cover all the variety~\cite{villar_etal2019}.
Moreover, non-Ia SNe are even more diverse, while typical templates for each class are based only on few specific objects~\cite{sncosmo}.
Thus, the performance of algorithms, trained and tested on such a dataset, can be overestimated.
That is why we choose to use only real photometric observation data in this paper.

We aim to create an automatic classification algorithm of SNIa, based on it's transient light curve alone, without any host or redshift information. To train the algorithm, we use photometric data and object labels from the Open Supernova Catalog (OSC)\footnote{\url{http://sne.space}}~\cite{Guillochon_etal2017}, which includes almost all publicly available observations of supernovae, supernova candidates and objects which were previously wrongly classified as supernovae.
Even though OSC data is very heterogeneous and affected by observational selection (for example, transients that look like SNIa has more chances to have spectroscopic follow-up and be classified), we assume that classifier build upon real observation data can cover greater diversity of astrophysical objects.
The OSC has already been used for machine-learning classification problems. \citet{Muthukrishna_etal2019DASH} prepared supernova spectral classification to obtain type, age, redshift and other properties of the target object.
In~\citet{pruzhinskaya_etal2019,Ishida_etal2019} authors built an anomaly detection pipeline to find abnormal light curves in the OSC.
Method in~\citet{narayan_etal2018} profited both OSC and simulation photometric data to implement SN photometric classification for ANTARES LSST broker\footnote{\url{http://antares.noao.edu}}.
In the current paper, we use less strict control criteria in data selection comparing to \citet{narayan_etal2018} that leads to a greater number of objects (see Section~\ref{sec:data}) and do not use any simulated data which can have intrinsic systematics.

Previous papers use a wide range of methods to train classification algorithms on supernovae light curve data. In particular, in~\citet{2018MNRAS.473.3969R}, authors based their research on the dataset presented in SPCC~\citep{Kessler_etal2010}. This dataset includes 18,321 simulated SNe. Each object is described by multicolor light curves. Their models are trained using only candidates with the host galaxy redshift information and with not less than three observation points. This selection reduces their input data to 17330 objects. The authors suggest a 2-step classification approach training and optimising hyperparameters of the Diffusion Map and Random Forest Classifier (RFC) simultaneously. It allows the use of, not original light curve vectors, but vectors of similarity between objects as an input to RFC. This method shows a promising result with 0.96 ROC AUC.

Authors of~\citet{Boone2019} proposed an approach for SN~Ia classification of PLAsTiCC and achieved ROC AUC of 0.957 with a tree-based LightGBM model. Unlike our experiments, they use all available passbands and preprocess them with Gaussian process regression to smooth light curves. These GP models allow them to augment a part of the dataset with spectroscopically-confirmed objects.

Authors of~\citet{villar_etal2019,Villar_etal2020} classified transients of Pan-STARRS using machine learning.
In~\citet{villar_etal2019} only real data were used with machine learning classifiers applied to features extracted from a parametric fit of light curves.
The paper~\citet{Villar_etal2020} used a semi-supervised approach, with a neural network autoencoder trained on simulated data and its latent layer used as a feature vector for a classifier.

In~\citet{Moller_Boissi2020}, the authors consider the classification of 1,983,213 simulated SNe light curves. The number of candidates makes it possible to use complex models, such as recurrent and convolutional neural networks (RNN and CNN). CNN classification allows us to achieve excellent quality with 0.98 ROC AUC. However, they also tried to a Random Forest classifier (RF), and it showed an even better result with 0.9929 ROC AUC. This result indicates that RF works well with SNe data, and is suitable for solving our problem.

Two datasets from PLAsTiCC and Subaru/Hyper Suprime-Cam transient survey was used in~\citet{Takahashi_etal2020}. They had shown that artificial neural network classifier trained on PLAsTiCC data could be used to predict labels of real data, although, this leaded to a significant loss of accuracy. 

In this paper, we present a novel light curve feature extraction method while using well-known classification methods: logistic regression, random forest, gradient boosting and artificial neural network. Corresponding source code is publicly available\footnote{\url{https://github.com/HSE-LAMBDA/supernovae_classification}}.

The rest of the paper is organised as follows. In Section~\ref{sec:data}, we formally describe the problem and the data used for the analysis and their preprocessing. Section~\ref{sec:method} is devoted to feature extraction and machine-learning models. Results and their comparison to different machine learning approaches are shown in Sections~\ref{sec:res} and~\ref{sec:discuss} correspondingly. We conclude the paper in Section~\ref{sec:conclusion}.

\section{Problem Statement and Data sets\label{sec:data}}
This study aims to develop an approach for supernova classification, based only on the photometric information. Given the differences between data and simulation, the algorithm must be trained using as much information from available data as possible. In machine learning terms, we use a supervised approach that provides the most stable results.

OSC~\cite{Guillochon_etal2017} is a compilation of different catalogues and individual papers based on observations using various facilities and data processing pipelines.
This way of data collection makes OSC data very heterogeneous.
As of August 1st, 2020, OSC consisted of 70,964 objects\footnote{Our snapshot of OSC data can be found at \url{http://sai.snad.space/sne20200801/}}, 60,592 of which had photometric observations and only 9,985 had spectroscopic data.
To make our dataset as homogeneous as possible, we consider only light curves in the $r$-band. 
Data points marked as upper limits and those without recorded uncertainties are not used in this study since they can have a sizeable systematic deviation from well-measured data of large surveys. We use fluxes whenever they are given by OSC. We also convert magnitudes to fluxes assuming that the same photometric system was used for all observations of an individual object. The time is not corrected to the rest frame as redshift is not known for all objects. This selection produces 8,657 objects described by the following fields: name, object type, number of observations, timestamps of observation points, $r$-fluxes for each observation and corresponding flux uncertainties. We follow OSC recommendations for multiple-labelled objects and use the first label as a true one.
We labelled Type Ia SNe as Ia, and all other types, including peculiar Type Ia SN, were labelled as non-Ia.

\subsection{Filtering Data}

Often the quality of observations is not good enough and it makes impossible to extract information about the type of object. For example, there are too few observations or too much noise in SN light curve. Therefore, we develop a method for filtering the initial dataset, to avoid such objects during training.

\begin{figure}[!h]
\centering
\includegraphics[width=1.\linewidth]{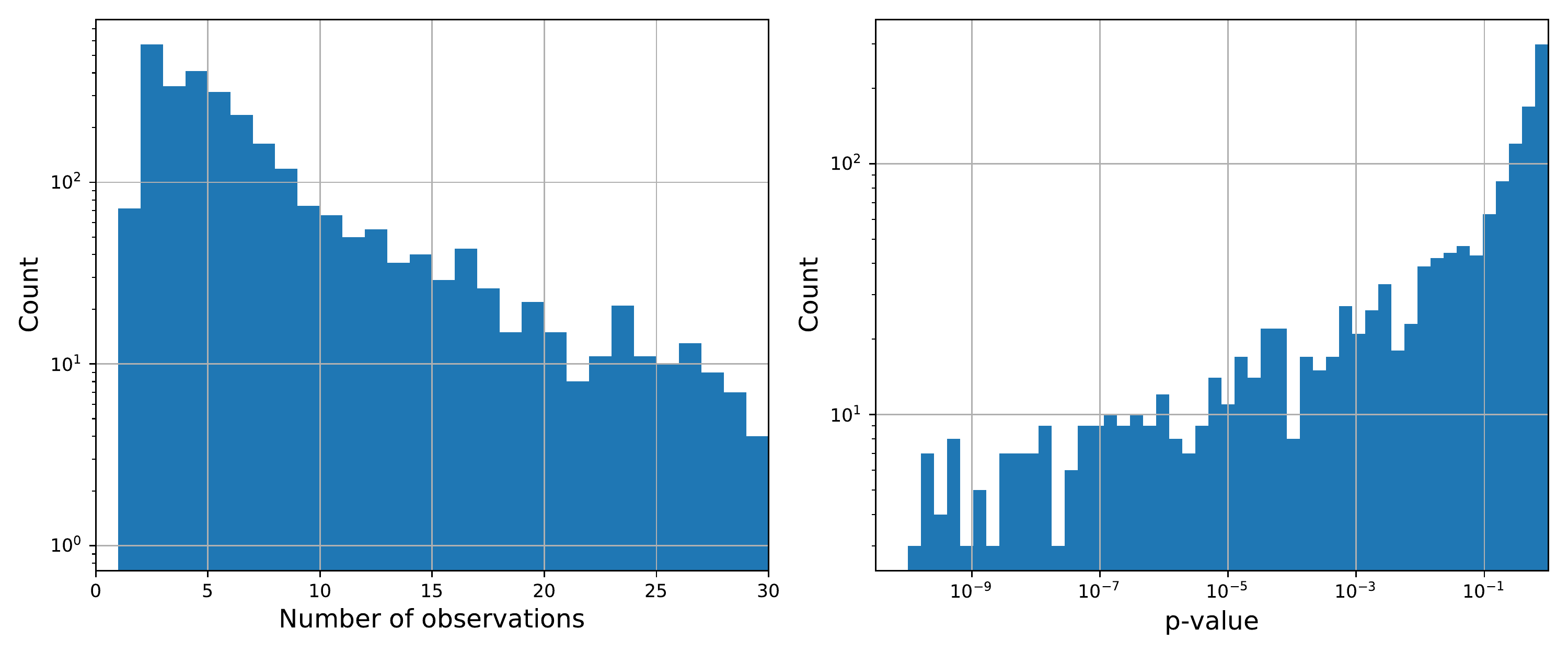}
\caption{Histogram of observations and a p-value of $\chi^2$-distributions for the candidates under consideration}
\label{fig:pval_dist}
\end{figure}

Consider a light curve with $n$ observations:

\begin{equation}
F(t_1), F(t_2), ..., F(t_{n-1}), F(t_n),
\end{equation}
where $F(t_i)$ is a light flux measurement at time $t_i$. We suggest two criteria for filtering the source data: number of light curve observations for each object and a $p$-value of $\chi^2$ per degree of freedom:
\begin{equation}
\bar{F} = \frac{1}{n} \sum_{i=1}^{n} F(t_i),
\end{equation}

\begin{equation}
\chi^2 = \frac{1}{n-1}\sum_{i=1}^{n} \frac{(F(t_i)-\bar{F})^2}{\sigma(t_i)^2},
\label{eq:rms}
\end{equation}

\begin{equation}
p(\chi^2, n-1) = 1 - \int_{0}^{\chi^2} \frac{z^{(n-3)/2} e^{-z/2}}{2^{(n-1)/2} \Gamma ((n-1)/2)} dz,
\label{eq:chi2}
\end{equation}
where we have $n-1$ degrees of freedom and $\Gamma (\cdot)$ is the Gamma function. Here, a $p$-value is calculated for the null hypothesis $H_0$: light curve is a constant function. Alternative hypothesis $H_1$ is that the curve has any other shape. If the $p$-value is smaller than a chosen significance level $\alpha$, the null hypothesis $H_0$ is rejected. Figure \ref{fig:pval_dist} shows a distribution of number of points on the left-hand panel and a distribution of $p$-value on the right-hand panel.

\begin{figure}[!h]
\centering
\includegraphics[width=1\linewidth]{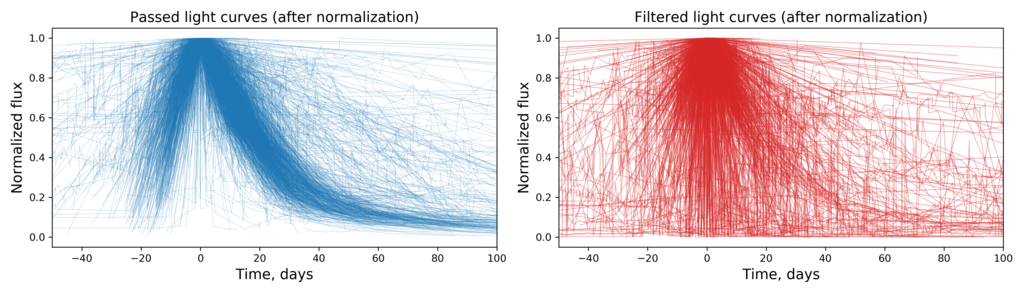}
\caption{Light curves that pass selection (blue curves) to the rest of candidates (red curves).}
\label{fig:cut_curves}
\end{figure}

We select supernovae candidates which are described with more than three observation points. Using a selection on $p$-value, we veto the candidates that do not look like transients because their light curves cannot be distinguished from constant function. We thus concentrate only on those objects with a $p$-value less than the significance level $\alpha = 0.001$. Applying the selection above, we obtain 1,527 objects, which we use to train our classification model. This dataset contains 1,184 SNe type Ia and 344 non-Ia. The latter includes 183 core-collapse SNe; 37 peculiar Type Ia SNe; 25 unconfirmed SN candidates with unknown true type; 44 super-luminous SNe and 54 misclassified objects of various true types, the largest of which is cataclysmic variables containing 31 objects.
The fraction of Type Ia SNe is close to limited volume surveys, such as ZTF Bright Transient Survey, which spectroscopic classified 73\% of transients as Type Ia SN\citep{ztf_bright2020}.
Figure~\ref{fig:cut_curves} shows the difference between filtered light curves and those that do not pass our selection. We observe that for rejected candidates, either observation points of filtered objects are actually in a shorter time range or light curves are too flat.

\section{Methods\label{sec:method}}

\subsection{Overview}\label{sec:methods_overview}

Light curves characterize the physical properties of observed astronomical objects. Their shape can be used to identify the type of objects, including different types of supernovae. The curves demonstrate a variety of shapes with lots of peculiarities typical for various types. While recent papers~\cite{pruzhinskaya_etal2019,Boone2019} propose to use augmentation techniques, we argue that the light curve peculiarities may be lost, or undesirable artefacts may appear after applying augmentation methods. That is why to keep the purity of the experiment, we use the curves with minimal transformations without additional augmentations applied. Each light curve has a different number of observation points, thus making it hard to input them directly into many classification algorithms. Our goal is, therefore, to describe all curves with the same number of observables, set up the same time reference point and bring the observations to the same scale of the light flux. These are the minimal steps required before applying machine learning models for the light curves classification. To check our strategy, we also perform a cross-check with augmentation of light curves dataset using Gaussian processes in~\ref{sec:discuss}.

After filtering out bad candidates, we proceed to the light curves preprocessing. We follow three preprocessing steps and a feature extraction approach which allows us to emphasise the physics properties of SN light curves. The steps are described in Subsection~\ref{sec:feature}. Finally, we train the models as it is described in Section~\ref{sec:model}.

\subsection{Light Curve Preprocessing\label{sec:light_curve_processing}}
\begin{figure}[h]
\centering
\includegraphics[width=1.\linewidth]{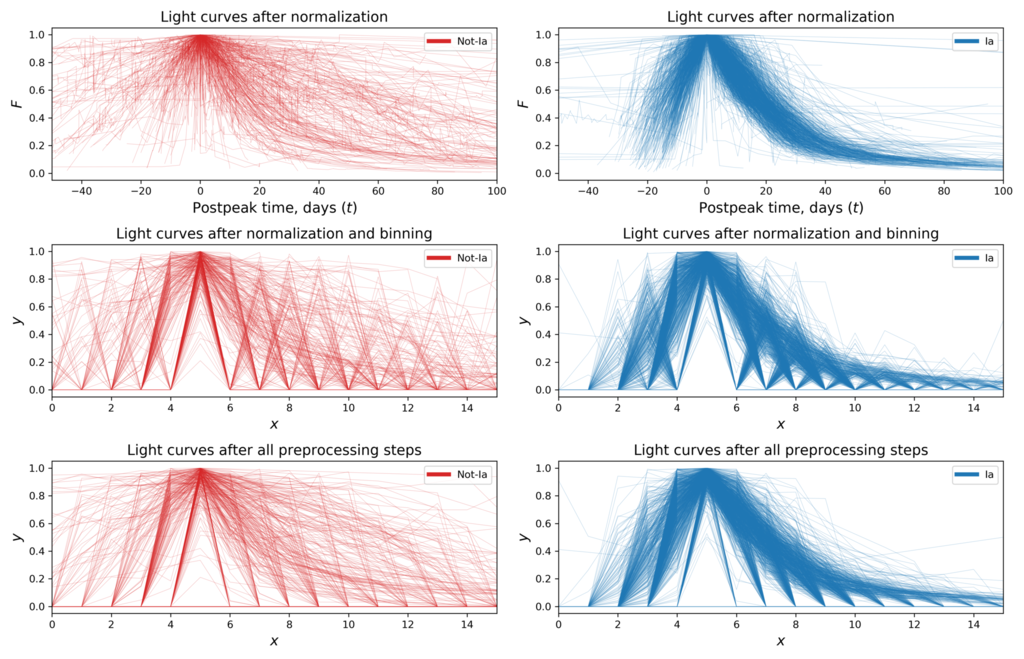}
\caption{Light curves from sources identified as type Ia (right-hand panels) and non-Ia (left-hand panels) after the Normalization and shift (top row), binning (middle row), and  linear interpolation (bottom row) }
\label{fig:preprocessing1}
\end{figure}

Initially, light curves are presented as arrays of photometric observations. To make comparable vector representations of objects, we suggest three steps: normalisation, binning and interpolation. Figure~\ref{fig:preprocessing1} shows us the intermediate results of preprocessing.

Normalisation consists of two steps: time shifting and light flux scaling. The observation times are shifted relative to the peak time $t_\mathrm{peak}$:
\begin{equation}
    t_\mathrm{peak} = \arg \max_{t_i} F(t_i)
\end{equation}
\begin{equation}
    t_i' = t_i - t_\mathrm{peak}
\label{eq:shifting}
\end{equation}

The light flux is scaled so that the highest value is 1:

\begin{equation}
    F'(t_i) = \frac{F(t_i)}{\max_{t_i} F(t_i)},
\label{eq:normalization}
\end{equation}

Binning allows us to get vectors of the same length and to consider them in the same vector space.
We divide the initial light curve in range from $t_\mathrm{left}'=-50$ to $t_\mathrm{right}'=100$ into $m=16$ equal bins (segments) with calculating the mean value for each bin:
\begin{equation}
      x_i = i ,~ i=0\,..\,m-1,
\end{equation}

\begin{equation}
      \tau_i = t_\mathrm{left}' + \frac{t_\mathrm{right}' - t_\mathrm{left}'}{m} x_i ,
\label{eq:t_for_bin}
\end{equation}

\begin{equation}
      y_i = \left\{
      \begin{aligned}
            &\frac1{n_i} \sum_{t' \in [\tau_i, \tau_{i+1})}{F'(t')}, &n_i \neq 0,\\
            &0, &n_i = 0.
      \end{aligned}
      \right.
\label{eq:bin}
\end{equation}
where $[t_i, t_{i+1})$ is the $i$-th bin interval, $y_i$ is a value of $i$-th bin, $n_i$ is a number of observations within a bin.

We select this number of bins based on the distribution of the number of observations in light curves, which is demonstrated in Figure~\ref{fig:pval_dist}. It shows that there are no curves with more than 30 observations, and most of them have less than 16 observations. Using much more bins does not provide more information about the objects but significantly increases the dimensionality of the input feature space, which may lead to the classification quality reduction.
Moreover, we assume that typical variability of SNIa on a timescale of 10 days should be larger than systematic uncertainties of photometric data which comes from differences in photometric systems of different observation facilities and can be as high as 20\% of the flux, so averaging along such a wide bin can reduce the influence of such systematical effects.
Our experiments show that taking more bins demonstrates the same or worse quality.

Some bins do not have any observation points, thus having zeroes in the corresponding places of the resulting vector. Intuitively, a supernova cannot dim completely for a moment and then light up again, but due to a different frequency of observations for each object, some experimental points are missing. That is why we use linear interpolation to fill missed observations. For example, one can consider a light curve, where $y_i \ne 0$ and  $y_{i+k} \ne 0$, but $y_{j} = 0$ for $i < j < i+k$. Then, the linear interpolation for $y_{j}$ is defined as follows:

\begin{equation}
    b_1 = \frac{y_{i+k} - y_{i}}{x_{i+k} - x_{i}},
\end{equation}
\begin{equation}
    b_0 = y_{i+k} - b_1 x_{i+k},
\end{equation}
\begin{equation}
    y_j = b_1 x_{j} + b_0.
\end{equation}

The resulting light curves are shown in Figure~\ref{fig:preprocessing1}.

\subsection{Feature Engineering\label{sec:feature}}
\begin{figure}[h]
\includegraphics[width=1.\linewidth]{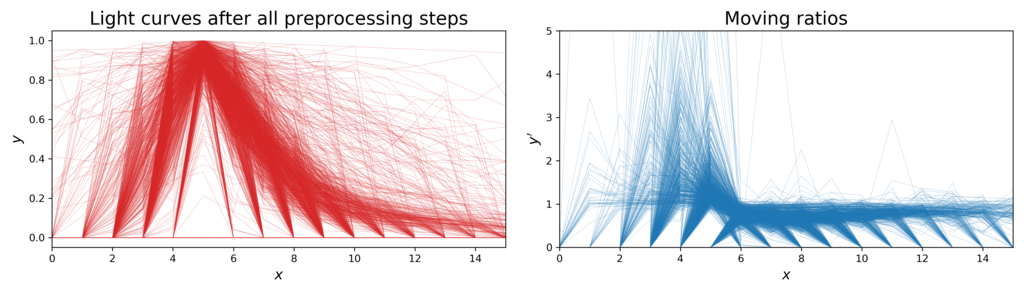}
\caption{Light curves for all Ia and Not-Ia SN objects in our data after all preprocessing steps.}
\label{fig:feature_gen}
\end{figure}

As noted earlier, an essential difference between Ia and non-Ia supernova is the rise-time and decline-time of their brightness. In order to emphasise the physics nature of supernovae light curves and improve the quality of the model, we generate an additional 16 features: for all elements $y_i$ of the resulting vector, we take the moving ratio of its elements:
\begin{equation}
y'_i = y_{i+1} / y_{i}.
\label{eq:relation}
\end{equation}
The resulting light curves are shown in Figure~\ref{fig:feature_gen}.

Finally, we concatenate the resulting vector with the one obtained as a result of the processing steps:
\begin{equation}
[y_0, y_1, ..., y_{m-1}, y_0', y_1', ..., y_{m-1}']^{T}
\end{equation}

\subsection{Model\label{sec:model}}

In our paper we consider four classification algorithms based on statistical models: Logistic Regression (LogReg), Random Forest Classifier~\cite{breiman1999random} (RF) implemented in the Scikit-learn python library, Gradient Boosting~\cite{friedman2001greedy} (XGBoost) implemented in the XGBoost python library and One-Layer Neural Network (One-Layer NN) using Keras library.

LogReg model is used with $L2$ regularization with constant $C=1.0$. RF model consists of 1000 decision trees with the maximum depth of 10. Additionally, the following values of hyperparameters are used: $\textit{max\_features} = \text{'auto'}$, $\textit{min\_samples\_split} = 10$ and default values for others. XGBoost model also has 1000 decision trees with the maximum depth of 6. Other values of hyperparameters are $\textit{learning\_rate} = 0.02$, $\textit{subsample} = 0.8$, $\textit{colsample} = 0.7$ and defaults for the rest. One-layer NN contains 32 neurons with ReLU activation function in the hidden layer and 1 output neuron with Sigmoid activation. The model is fitted to minimize a binary cross-entropy loss function using Adam optimizer during 50 epochs with $\textit{learning\_rate} = 0.001$. 

To train any model, we need to split our data into two subsets: one for training and one for testing. In this case, we lose a significant part of our data. To solve this problem, we use a K-Fold cross-validation approach with $k = 10$.

\section{Results\label{sec:res}}

After training the models, we predict labels for all objects in the initial dataset, so we get a true label and a predicted label for each SN object. To evaluate our model performance we use five common metrics: area under the receiver operating characteristic curve (ROC AUC), accuracy, $F_1$-score, precision and recall. In our paper, we consider ROC AUC as a target metric as it takes into account both true positive and false positive rates. Metric scores for each model are presented in Table \ref{table:osc}.

\begin{table}[!h]
\centering
\begin{tabular}{||c c c c c||} 
 \hline
 Metric/Model & LogReg & RF & XGBoost & One-Layer NN \\ [0.5ex] 
 \hline\hline
 ROC AUC & $0.801\pm0.016$ & $\textbf{0.876}\pm0.011$ & $0.866\pm0.012$ & $0.867\pm0.011$ \\ 
 \hline
 Accuracy & $0.861\pm0.009$ & $\textbf{0.864}\pm0.009$ & $0.859\pm0.009$ & $0.859\pm0.009$ \\
 \hline
 F1-score & $0.916\pm0.006$ & $\textbf{0.917}\pm0.006$ & $0.912\pm0.006$ & $0.914\pm0.006$ \\
 \hline
 Precision & $0.860\pm0.010$ & $0.874\pm0.009$ & $\textbf{0.882}\pm0.009$ & $0.866\pm0.009$ \\
 \hline
 Recall & $\textbf{0.979}\pm0.004$ & $0.964\pm0.006$ & $0.943\pm0.007$ & $0.969\pm0.005$ \\ [1ex] 
 \hline
\end{tabular}
\caption{Metrics}
\label{table:osc}
\end{table}

We use the Logistic Regression model as one of the simplest classifiers to estimate the lowest classification quality can be archived on our data. LogReg shows good results, which means that presented objects are well separated and can be efficiently classified with linear models. RF classifier shows the best scores in all metrics except precision and recall, which can be explained by the fact that the model is resistant to outliers. We choose it as our primary model, and all feature analysis is based on RF model results.

\begin{figure}[!h]
\centering
\includegraphics[width=1.\linewidth]{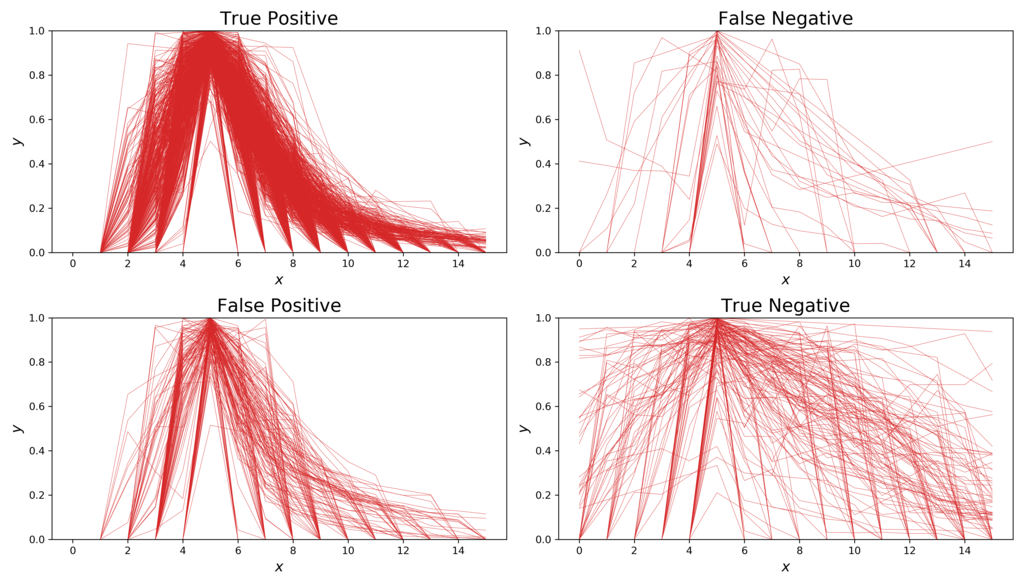}
\caption{Preprocessed Light Curves by type of error}
\label{fig:confusion}
\end{figure}

\begin{figure}[!h]
\centering
\includegraphics[width=1.\linewidth]{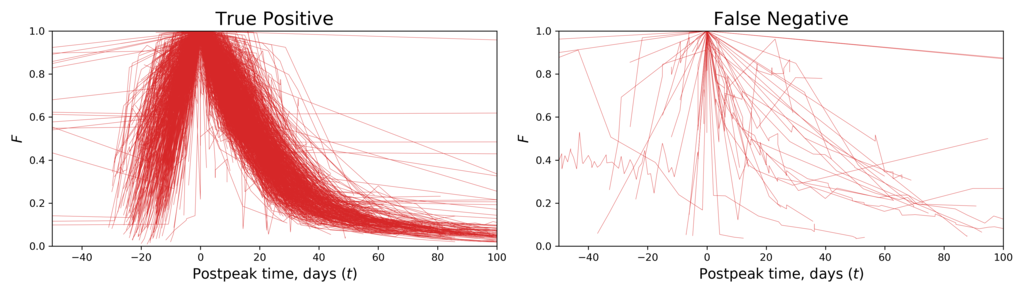}
\caption{Comparison of TP and FN objects}
\label{fig:init_data_confusion}
\end{figure}

To evaluate the quality of our model more thoroughly, we calculate a confusion matrix for the RF classifier predictions as shown in Table~\ref{table:confus}. For example, a cell in the first row and the second column of the table represents the number of objects that are Ia, but predicted as Not-Ia (negative). This number is called False Negative (FN). Light curves of objects in FN subset in Figure~\ref{fig:confusion} are more noisy than curves in the True Positive (TP) subset. We conclude that in this case, we are dealing with a bias of initial data quality. If we consider FN light curves in Figure~\ref{fig:init_data_confusion}, we see that most of them have poor quality even though they passed our filter.

\begin{table}[h]
\centering
\begin{tabular}{l|l|c|c|c}
\multicolumn{2}{c}{}&\multicolumn{2}{c}{Prediction}&\\
\cline{3-4}
\multicolumn{2}{c|}{}& Ia (positive) & Not-Ia (negative) &\multicolumn{1}{c}{Total}\\
\cline{2-4}
\multirow{2}{*}{Actual}& Ia & $1141$ & $43$ & $1184$\\
\cline{2-4}
& Not-Ia & $164$ & $179$ & $343$\\
\cline{2-4}
\multicolumn{1}{c}{} & \multicolumn{1}{c}{Total} & \multicolumn{1}{c}{$1305$} & \multicolumn{    1}{c}{$222$} & \multicolumn{1}{c}{$1527$}\\
\end{tabular}
\caption{Confusion Matrix for the Random Forest classifier predictions. An object is predicted as Ia SN if the classifier output is larger than $0.5$. Otherwise, it is considered as Not-Ia SN objects.}
\label{table:confus}
\end{table}

\begin{figure}[!h]
\centering
\includegraphics[width=0.8\linewidth]{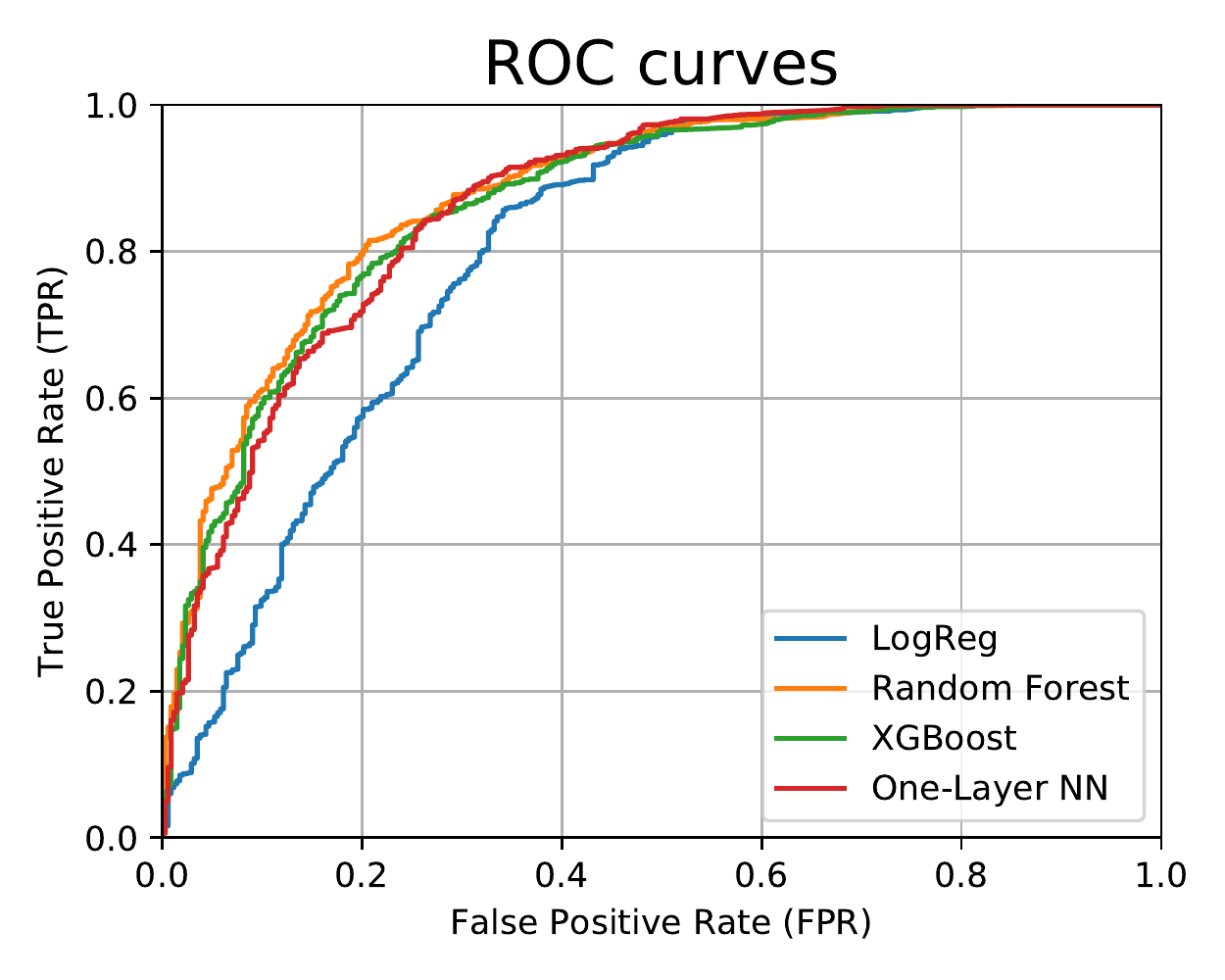}
\caption{ROC Curves for all models}
\label{fig:roc_curves}
\end{figure}

\begin{figure}[!h]
\centering
\includegraphics[width=1\linewidth]{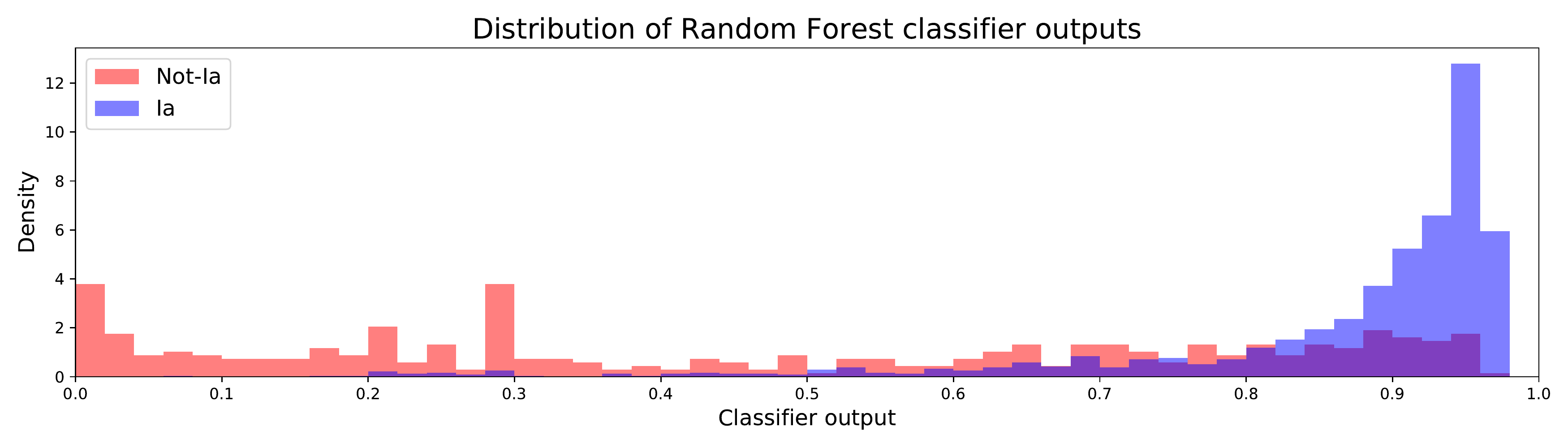}
\caption{Distribution of the Random Forest classifier output for true Ia (blue) and Not-Ia (red) objects.}
\label{fig:prob_dist}
\end{figure}

The ROC curves are demonstrated in Figure~\ref{fig:roc_curves}. The histogram of two classes colored by blue (Ia) and red (non-Ia) with respect to RF model output demonstrated in Figure~\ref{fig:prob_dist}. It shows us that a huge part of supernovae can be correctly classified with high probabilities.

\begin{figure}[h]
\centering
\includegraphics[width=1\linewidth]{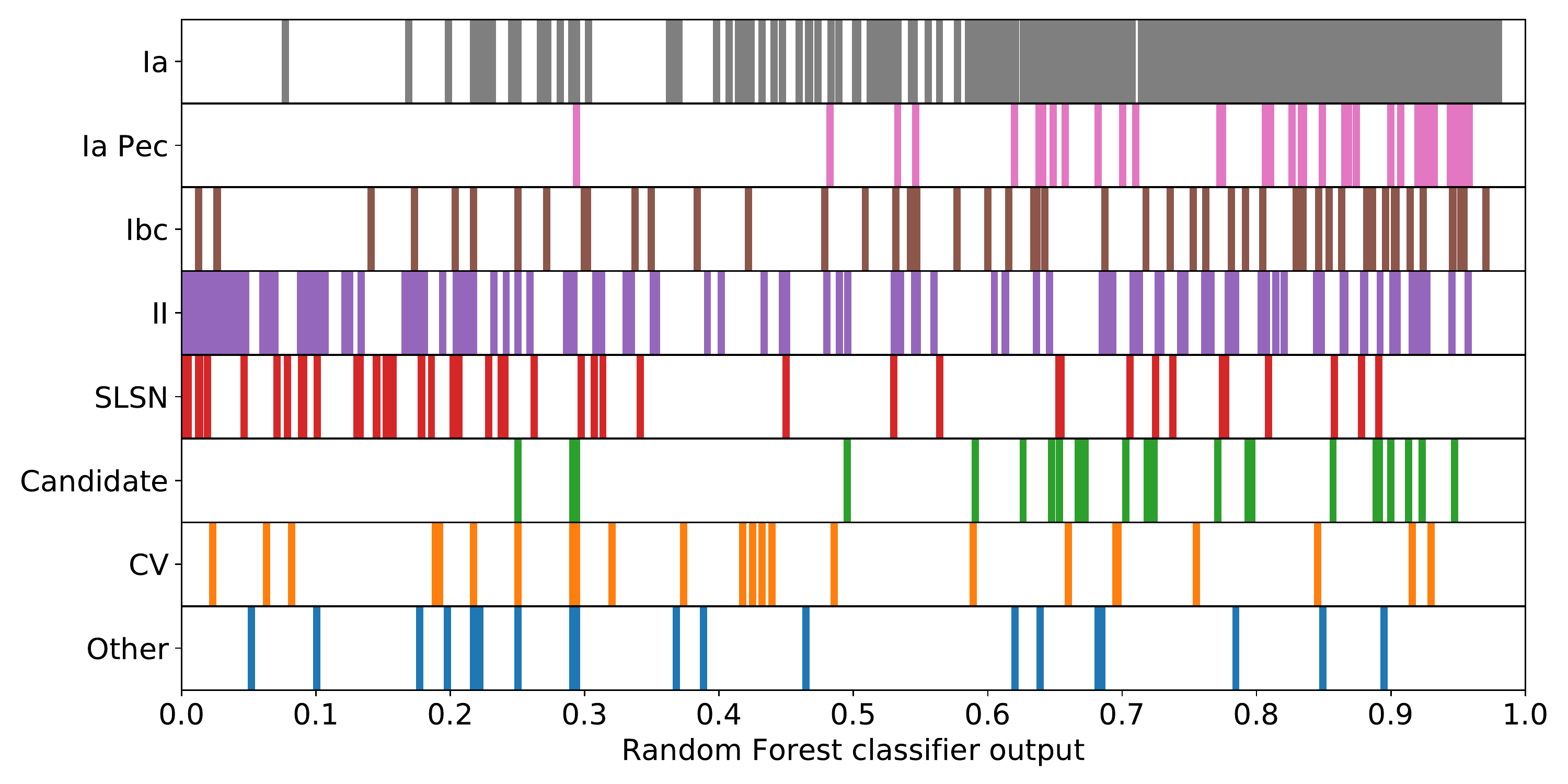}
\caption{The Random Forest classifier outputs for different transient types.}
\label{fig:pred_for_types}
\end{figure}

Figure~\ref{fig:pred_for_types} shows output of the Random Forest classifier for different types of objects.
Light curves of peculiar Type Ia SNe are very similar to ordinal SNe Ia, so the classifier cannot tell them apart.
Roughly the two thirds of Type Ib and Ic SNe were labeled as Ia, but we evaluate this result as an acceptable compared to papers on multi-class transient classification~\cite{Boone2019,Takahashi_etal2020,Muthukrishna_etal2019}.
However the quality of Type II and super-luminous SNe classification is acceptable.
The objects marked as candidates in OSC are likely to be unconfirmed Type Ia SN, so the classifier labeled almost all of them as Ia.
The majority of cataclysmic variables are labeled as non-Ia.

\section{Discussion\label{sec:discuss}}

In this article, we suggest an SN classification method in which we use only real photometric data for model training. Previous works described in Section~\ref{sec:intro} have shown promising results using samples of simulated SNe. Nevertheless, we suggest to use with great care, any data simulations during model training for real objects classification, as such models can overfit to simulated samples and show less precise classification results for real physical objects observed with future experiments. To demonstrate the statement above we train two RF models on simulated and real datasets: PLAsTiCC and OSC respectively.

In our case, we use the same preprocessing pipeline and feature extraction methods for both datasets as described in Subsection~\ref{sec:light_curve_processing}. This gave us 2605 light curves with the similar redshift and number of observation distribution to OSC. Then we test it with PLAsTiCC and OSC test datasets. As a result, we get four train-test combinations: RF trained with OSC and tested with PLAsTiCC, RF trained with PLAsTiCC and tested with OSC, RF trained and tested with OSC and RF trained and tested with PLAsTiCC. In order to make the models resulting scores as comparable as possible, we undersample non-Ia class in the PLAsTiCC dataset to achieve OSC dataset class balance. We also select only 8 types from the PLAsTiCC dataset \citep{PLAsTiCC-models} which are present in OSC: SNIax , SNIa, SNIa-91bg, SNII (which includes both IIP/L and IIn), SNIbc, SLSN-I, Tidal Disruption Event (TDE), and Active Galaxy Nucleus (AGN).

\begin{figure}[!h]
\centering
\includegraphics[scale=0.8]{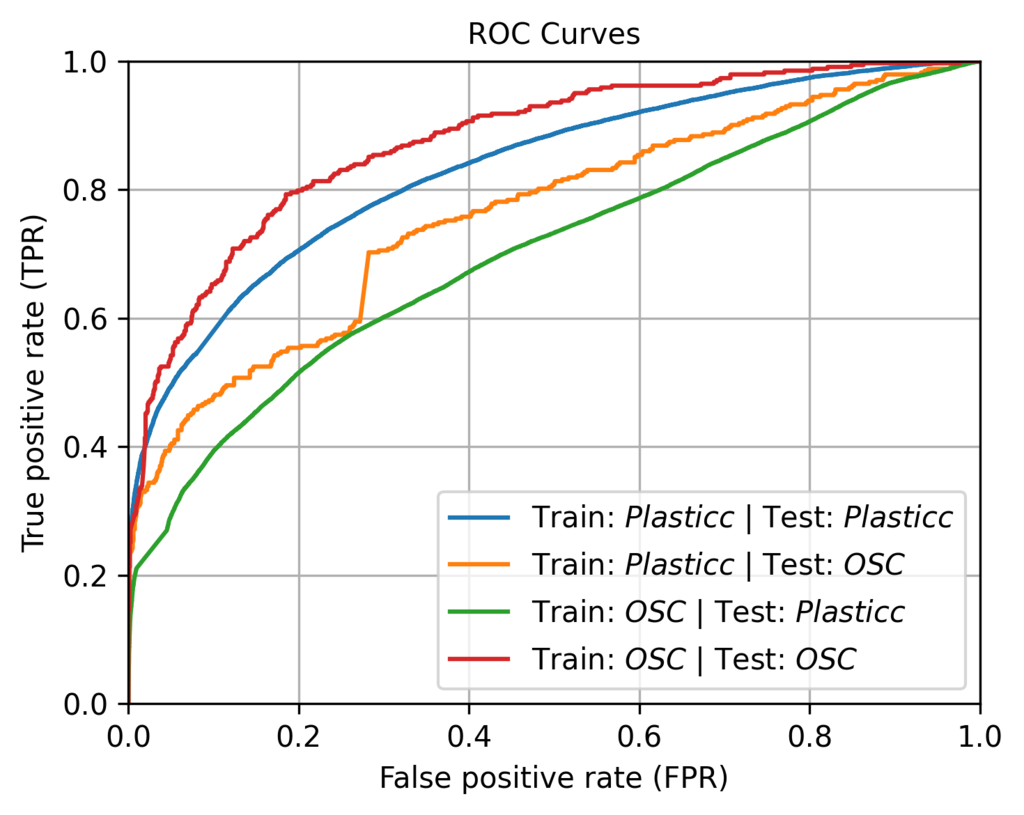}
\caption{ROC Curves for different train-test combinations}
\label{fig:roc_plasticc}
\end{figure}

Table~\ref{table:osc-plasticc} shows that the performance of a model trained with simulated data decreases its ROC AUC score, from 0.843 to 0.739, while testing on real objects. The same effect is observed in the inverse experiment. The corresponding ROCs are shown in Figure~\ref{fig:roc_plasticc}. However, learning transfer from the PLAsTiCC to the real data might work better for one-survey light curves, such as ZTF or Pan-STARRS.

\begin{table}[!h]
\centering
\begin{tabular}{||c c c c c||} 
 \hline
 Metric/Model & PLAsTiCC-PLAsTiCC & PLAsTiCC-OSC & OSC-PLAsTiCC & OSC-OSC \\ [0.5ex] 
 \hline\hline
 ROC AUC & $0.833\pm0.002$ & $0.761\pm0.016$ & $0.702\pm0.001$ & $0.876\pm0.011$ \\ 
 \hline
 Accuracy & $0.848\pm0.001$ & $0.816\pm0.010$ & $0.780\pm0.001$ & $0.864\pm0.009$ \\
 \hline
 F1-score & $0.908\pm0.001$ & $0.886\pm0.007$ & $0.862\pm0.001$ & $0.917\pm0.006$ \\
 \hline
 Precision & $0.850\pm0.001$ & $0.855\pm0.010$ & $0.835\pm0.001$ & $0.874\pm0.009$ \\
 \hline
 Recall & $0.974\pm0.001$ & $0.918\pm0.008$ & $0.891\pm0.001$ & $0.964\pm0.005$ \\ [1ex] 
 \hline
\end{tabular}
\caption{Metrics obtained for different train-test combinations}
\label{table:osc-plasticc}
\end{table}

While some of the papers mentioned above show better results, in practice, they suffer from an unknown systematic uncertainty due to the simulated events sample used in training. Our approach is free of this virtue and shows that using only a small amount of real data, one can obtain a scalable result, which might be improved once more data is available. 

A popular approach to augment light curve samples is to use Gaussian Processes (GP) and then describe them by a set of features~\cite{Boone2019, pruzhinskaya_etal2019}. To test this on the OSC data, we use GP to approximate the light curves. We use the sum of radial basis function (RBF) and white noise kernels for the covariance matrix calculation and take into account observation errors of the curves. Then, we repeat the preprocessing steps described in Subsection~\ref{sec:light_curve_processing} with a various number of bins. The results show that GP does not significantly change the classification quality.

Then, we describe the GP approximations of the light curves by a set of features provided in~\citet{cabral2018fats} and use them as inputs of the classification models. The experiments demonstrate that the classification quality is worse than shown in this paper. As we mentioned in Subsection~\ref{sec:methods_overview}, the augmentation techniques do not help to significantly improve the quality of Ia type supernovae classification on the OSC data, so we do not use them in this work.

\section{Conclusion\label{sec:conclusion}}
In this paper, we consider a new data-driven classification approach of SN objects from the Open Supernovae Catalog. We managed to achieve good classification quality using only real objects. To emphasise the physical properties of SN Ia, we suggest adding a vector of generated features to the initial vector. The random forest classifier shows the best performance, with the highest score. We demonstrated that training a model on the PLAsTiCC simulated data significantly reduces its efficiency in classifying real objects. Hence, we can conclude that validation of the model on real data is a necessary step for the purpose of achieving good classification quality on real-life tasks.

\section*{Acknowledgements}
The authors thank Maria Pruzhinskaya for the fruitful discussion.
KM is supported by a RFBR grant 20-02-00779 for preparing the Open Supernova Catalog AND PLASTiCC data. DD is supported by a RSF 19-71-30020 grant in preparing and discussing data augmentation techniques and data-simulation discrepancy measurement methods. This research has made use of NASA's Astrophysics Data System Bibliographic Services and following {\sc Python} software packages: {\sc NumPy}~\citep{numpy}, {\sc Matplotlib}~\citep{matplotlib}, {\sc SciPy}~\citep{scipy}, {\sc pandas}~\citep{pandas}, and {\sc scikit-learn}~\citep{scikit-learn}.

\bibliographystyle{model1-num-names}
\bibliography{sample.bib}

\end{document}